\documentclass[aps,showpacs,twocolumn,floatfix,superscriptaddress]{revtex4}

\usepackage{graphicx}
\usepackage{amsmath}
\usepackage{amsfonts}
\usepackage{amssymb}
\usepackage{bm}

\begin{document}

\title{An electronic Mach-Zehnder interferometer in the Fractional Quantum Hall effect}

\author{T. Jonckheere},
\affiliation{ Centre de Physique Th\'eorique, Case 907 Luminy, 13288 Marseille cedex 9, France}
\author{P. Devillard}
\affiliation{ Centre de Physique Th\'eorique, Case 907 Luminy, 13288 Marseille cedex 9, France}
\affiliation{Universit\'e de Provence, 13331 Marseille cedex 3, France}
\author{A. Cr\'epieux}
\affiliation{ Centre de Physique Th\'eorique, Case 907 Luminy, 13288 Marseille cedex 9, France}
\affiliation{Universit\'e de la M\'edit\'erann\'ee, 13288 Marseille cedex 9, France}
\author{T. Martin}
\affiliation{ Centre de Physique Th\'eorique, Case 907 Luminy, 13288 Marseille cedex 9, France}
\affiliation{Universit\'e de la M\'edit\'erann\'ee, 13288 Marseille cedex 9, France}

\date{\today}

\begin{abstract}
We compute the interference pattern of a Mach-Zehnder interferometer operating in the 
fractional quantum Hall effect. Our theoretical proposal is inspired by 
a remarkable experiment on edge states in the Integer Quantum Hall effect 
(IQHE)~\cite{Ji-Chung.2003}. The Luttinger liquid model is solved via two 
independent methods: refermionization at $\nu=1/2$ and the Bethe Ansatz solution
available for Laughlin fractions. The current differs strongly  from that of 
single electrons in the strong backscattering regime. The Fano factor is periodic
in the flux, and at $\nu=1/2$ it exhibits a sharp transition from sub-Poissonian (charge $e/2$)
to Poissonian (charge $e$) in the neighborhood of destructive interferences.
Implications for Laughlin fractions are discussed. 
\end{abstract}


\maketitle

A fascinating aspect of mesoscopic physics is to build
 analogs of optical devices with the help of nanostructures.
In many situations both phenomena can be understood with the same language 
\cite{akkermans_montambaux}. However, photons propagate in vacuum and therefore
interact weakly, except during their generation/detection processes. 
On the opposite, interactions between electrons are manifest in one 
dimensional systems as well as in quantum dots.  
Here we want to inquire how electronic interactions affect the interference 
pattern of a classic optical device analog, a Mach-Zehnder (MZ) interferometer 
\cite{Born-Wolf.1999}. 

Recently, such an analog was 
achieved with edge states of the integral quantum Hall effect (IQHE) \cite{Ji-Chung.2003}.
Interference visibilities as high as $\sim 60\%$ were observed.
Edge states of the IQHE can be understood in principle at the 
single electron level, but at higher magnetic fields 
electronic interactions are explicit in the
fractional quantum Hall effect (FQHE). The latter offers
the opportunity to investigate fractional charge~\cite{Kane-Fisher.1994,Saminadayar-Glattli.1997} 
and fractional statistics \cite{laughlin} in one dimension.
Interferometry in the FQHE was previously 
studied with regard to fractional charge detection 
\cite{Chamon-Freed.1997} using perturbation theory. 
Here we report on MZ interferometry using exact 
models: refermionization at $\nu=1/2$ \cite{Chamon-Freed.1996} and the Bethe Ansatz solution  
\cite{Fendley-Ludwig.1995}. In the strong backscattering regime, the interference pattern
displays a dramatic effect of the interactions: 
the signal is not sinusoidal, and its amplitude at the output
departs from the single electron expectations.  

\begin{figure}
\centerline{\includegraphics[width=7.cm]{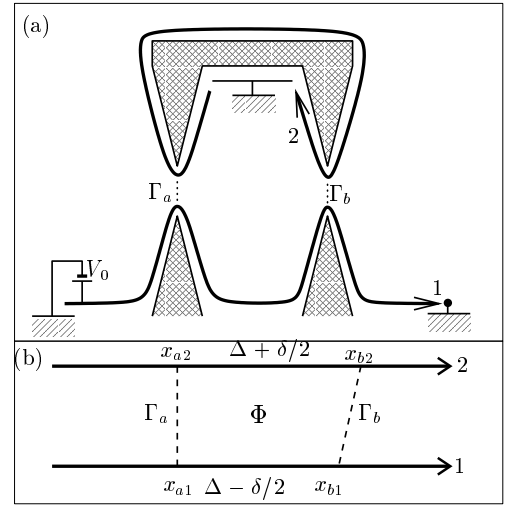}}
\caption{(a) Mach-Zehnder geometry in the quantum Hall effect: counter-propagating edge states
at a QPC are made to meet again at a second QPC. 
(b) Edge state configuration equivalent to (a). $\Gamma_a$  ($\Gamma_b$) is the tunneling amplitude at the
first (second) QPC, at $x=x_{a1}$ and $x=x_{a2}$  ($x=x_{b1}$ and $x=x_{b2}$) for edge state 1 (2).
The mean distance between the two QPCs is $\Delta$ and the path difference is $\delta$. $\Phi$ is the AB phase
due to the magnetic flux.}
\label{fig:setup}
\end{figure}

The MZ setup~\cite{Ji-Chung.2003} is depicted in Fig.~\ref{fig:setup} 
(a): an edge state is injected at voltage $V_0$, meets a quantum point contact 
(QPC) where it is scattered.
The two resultant states recombine at a second QPC, giving two outgoing edge states 
1 and 2. A magnetic field $B$ threads the surface $S$ enclosed by the 2 edges between the 2 QPCs, leading 
to an Aharonov-Bohm 
(AB) flux and to a corresponding phase $\Phi= S B/\Phi_0^*$, where $\Phi_0^* =  h c / e^*$ is
the flux quantum for excitations with fractional charge $e^* = \nu e$ ($\nu$ is the filling factor).
This setup is topologically equivalent to the one of Fig.~\ref{fig:setup} (b), 2 chiral states propagating 
in the same direction meeting successively two QPCs. 
This geometry is thus different from the simple Hall bar described 
in~\cite{Chamon-Freed.1997}.

We wish to calculate the outgoing current  in edge 1 and 2 for arbitrary tunneling amplitudes
$\Gamma_a$,$\Gamma_b$ -- thus non-perturbatively -- as a function
of the AB phase $\Phi$, the applied voltage $V_0= \hbar \omega_0 / e^*$, the mean distance 
$\Delta$ between the two QPCs and the path difference $\delta$.
The AB phase is a key ingredient to the problem, as it modulates 
the interferences between the two paths from the first
to the second QPC. As this phase is present only for cross terms with one tunneling event at 
QPC 1 and another at QPC 2, this amounts to multiplying the $\Gamma_a \Gamma_b^*$ terms by the phase $e^{i \Phi}$. 
The bosonized Hamiltonian reads (as in~\cite{Chamon-Freed.1996}, $\hbar=e=v_F=1$, 
except in important results):
\begin{multline}
\label{eq:bosonHamiltonian}
H=  H^0_{\phi_1} +  H^0_{\phi_2} \\ 
+ \sum_{q=a,b} \left( \Gamma_q e^{-i \omega_0 t} e^{i \sqrt{\nu} (\phi_1(x_{q},t)
 - \phi_2(x_{q},t)) } + \mbox{h.c.} \right)~. 
\end{multline}
The first two terms in $H$ are the free edge Hamiltonians, the next 
describe the tunneling of charge $\nu e$ through the two QPCs. 
We consider the case of equal distances between the two QPCs: the first (second) QPC is
located at $x_a$ ($x_b$) for both edge states;
unequal distances tend to reduce the interferences, but do 
not change the results  qualitatively. Note that, as our calculations will be nonperturbative,
the tunneling Hamiltonian is able to describe also electron tunneling at strong coupling, see~\cite{Chamon-Freed.1996}.

At $\nu=1/2$, it is natural to introduce new bosonic fields 
$
\phi_{\pm}(x)  =  (1/\sqrt{2}) \left( \phi_1(x) \pm \phi_2(x) \right)
$.
The tunneling operators in Eq.~(\ref{eq:bosonHamiltonian}) can be
represented by a fermionic field
$\eta(x) = e^{i \phi_{-}(x)}$.
The Hamiltonian contains a trivial 
free part for $\phi_{+}$, and a non-trivial 
part for $\phi_{-}$.
It is then possible to obtain a 
Hamiltonian which
is quadratic in fermionic variables, provided that new
fermionics fields are introduced such that 
$\psi(x,t) = \eta(x,t) \, f$, where $f$ is a Majorana fermion 
($f= C + C^{\dagger}$ and $\{ C,C^{\dagger} \} = 1$):
\begin{multline}
\label{eq:fermionpsiHamiltonian}
H_{-} = \int \! dx \bigg[ \psi^{\dagger}(x) \left (-i \partial_x - \omega_0 \right) \psi(x)\\
   + \sum_{q=a,b} \sqrt{2 \pi} \, \delta(x-x_q) \left (\Gamma_q \, \psi(x) \, f  + \Gamma_q^* \, f\, \psi^{\dagger}(x) \right) \bigg] 
~.\end{multline}
 $\psi(x)$ is propagating in ballistically, except at $x=x_a, x_b$. 
The Heisenberg equations for $\psi(x,t)$ are solved by introducing plane wave solutions:
$
\psi(x,t) = \sum_{\omega} u_{\omega} \, e^{i \omega_0 x} e^{i \omega (x-t)}
$,
with coefficients $u_{\omega} = A_{\omega}$ ($C_{\omega}$) for the incoming (outgoing) field
at the left (right) of the two QPCs. The boundary conditions at the QPCs give:
\begin{multline}
\label{eq:refermiosoluce}
C_{\omega} = D^{-1} \bigg[ \left(  i \omega - 4 \pi \widetilde{\Gamma}_a \widetilde{\Gamma}_b^* \, 2 i \, \sin(\omega \Delta) \right) A_{\omega}
\\ - 4 \pi \left( \left(\widetilde{\Gamma}_a^*\right)^2 +\left(\widetilde{\Gamma}_b^*\right)^2 +
                            2 \widetilde{\Gamma}_a^* \widetilde{\Gamma}_b^* \cos(\omega \Delta) \right)   A^{\dagger}_{-\omega} \bigg] ,
\end{multline}
with: 
$
D = i \omega - 4 \pi [ \left|\Gamma_a\right|^2 +  \left|\Gamma_b\right|^2 + 
                         ( \widetilde{\Gamma}_a \widetilde{\Gamma}_b^* + \widetilde{\Gamma}_a^* \widetilde{\Gamma}_b )
                                        e^{i \omega \Delta} ]
$ 
and the tunneling amplitudes are redefined as 
$
\widetilde{\Gamma}_{a,b} = \Gamma_{a,b} \, e^{i \omega_0 x_{a,b}} e^{\pm i \Phi/2}
$.
 Equation~(\ref{eq:refermiosoluce}) can be seen as 
the solution of a scattering problem.
Writing $C_{\omega} = r_{\omega} A_{\omega} + t_{\omega} A^{\dagger}_{-\omega}$, with reflection ($r_{\omega}$) and
transmission ($t_{\omega}$) coefficients, one can check that the flux is conserved ($|r_{\omega}|^2 + |t_{\omega}|^2 =1$).

 From the solution Eq.~(\ref{eq:refermiosoluce}), we can proceed to the calculation of the current $I_2$ outgoing in edge state  2:
\begin{equation}
\label{eq:I2S2}
I_2 = \frac{e}{4 \pi}  \, \int_{-\omega_0}^{\omega_0} \!\!\!\!\! d\omega \; 
   \left|t_{\omega}\right|^2 ~.
\end{equation}
The outgoing current in edge state 1 is simply  $I_1 = e \,\omega_0 / (2 \pi) - I_2 $, where 
$e \omega_0 / (2 \pi) = \nu e^2 V_0/h$ is the incoming Hall current.
It is convenient to introduce the geometric mean modulus amplitude $\Gamma=\sqrt{|\Gamma_a| |\Gamma_b|}$.
The deviation from equal amplitudes is described with the parameter $\lambda$
($|\Gamma_a| = \lambda \Gamma$, $|\Gamma_b| = (1/\lambda) \Gamma$). 
The transmission becomes:
\begin{eqnarray}
\left|t_{\omega}\right|^2 &=& N(u)/D(u)  \; , \quad  u=\omega/(4\pi \Gamma^2)
~,  \\
N(u) & = &  \left[ (\lambda^2 + \lambda^{-2}) \cos(\omega_0 \Delta + \Phi) + 2 \cos(4 \pi \Gamma^2 \, u \, \Delta) \right]^2 \nonumber \\
 & &  + \left[ (\lambda^2-\lambda^{-2}) \sin(\omega_0 \Delta + \Phi)\right]^2~, \nonumber \\
D(u) & = &  \left[ u - 2 \cos(\omega_0 \Delta + \Phi)  \sin(4 \pi \Gamma^2 \, u \, \Delta) \right]^2  \nonumber \\
 & &  + \left[ (\lambda^2+\lambda^{-2}) + 2  \cos(\omega_0 \Delta + \Phi)  \cos(4 \pi \Gamma^2 \, u \Delta) \right]^2~. \nonumber 
\end{eqnarray}
The relevant regimes for observing interference fringes are either weak pinchoff ($\Gamma \to 0$) or when $(\omega_0 \Delta/v_F) < 1$ at  strong pinchoff.
At strong pinchoff and for $(\omega_0 \Delta/v_F) \gg 1$, 
the above integral gives $e \,\omega_0/(4 \pi)$, and thus $I_1 = I_2 = e \, \omega_0 /(4 \pi)$, where all interferences
are lost. For $(\omega_0 \Delta/v_F) < 1$, the integral gives:
\begin{multline}
\label{eq:currentsoluce}
I_2 \simeq \left( 2 \Gamma^2 \right) \;  
 \frac{ \frac{1}{2}(\lambda^2 + 1/\lambda^2) + \cos(\omega_0 \Delta + \Phi) }{1/2 - 4 \pi \Gamma^2 \Delta \cos(\omega_0 \Delta + \Phi)} 
\\ \times 
\mbox{tan}^{-1}
\left(\omega_0  \frac{ \frac{1}{2}  -  4 \pi \Gamma^2 \Delta \cos(\omega_0 \Delta + \Phi)}
                            { 4 \pi \Gamma^2 \left(\frac{1}{2}(\lambda^2+ 1/\lambda^2) + \cos(\omega_0 \Delta + \Phi)\right)} \right)~.
\end{multline}    
Comparing this to the transmitted current for one QPC only, with tunneling amplitude $\Gamma$:
\begin{equation}
I_2|_{1 QPC} = 2 \Gamma^2 \, 
\mbox{tan}^{-1}
\left( \frac{\omega_0}{4 \pi \Gamma^2} \right)~,
\end{equation}
we see that the current in Eq.~(\ref{eq:currentsoluce}) can be expressed as the 
current for a single QPC, with an
effective tunneling amplitude $\Gamma_{eff}$:
\begin{equation}
\label{eq:gammaeff}
4 \pi \Gamma_{eff}^2 = \frac{4 \pi  \left( \frac{1}{2}(|\Gamma_a|^2 + |\Gamma_b|^2) + |\Gamma_a| |\Gamma_b| \cos\Phi\right)}
{1/2 - 4 \pi |\Gamma_a| |\Gamma_b| \, \Delta \cos\Phi}~.
\end{equation}
This is a central result: 
as far as the current is concerned, the MZ setup behaves, for fractionaly charged excitations,
as a single QPC with an effective amplitude $\Gamma_{eff}$ which is modulated by the AB phase. 
As,  in this setup which is composed of several edges, excitations are injected from one edge and are collected from another edge,
we expect that the fractional statitics of these excitations play an important role in this result (see~\cite{Safi-Martin.2001}).
Technically speaking, the difference between this behavior and the one of non-interacting electrons 
(as observed experimentally in the IQHE~\cite{Ji-Chung.2003}) can be traced back to the Hamiltonian. 
In the FQHE, the 
fermionic Hamiltonian of 
Eq.~(\ref{eq:fermionpsiHamiltonian}) couples $\psi(x,t)$ to the auxilliary fermion $f$
for both scattering events at
$x_a$ and at $x_b$: 
these two scatterings are thus strongly linked. 
On the other hand, in the 
IQHE, the tunneling part of the Hamiltonian
$
H_T= \int dx \; \sum_{q=a,b} \delta(x - x_q) \left( \Gamma_q \, \psi_1(x) \psi^{\dagger}_2(x) + \mbox{h.c.} \right)
$
couples $\psi_1(x,t)$ to $\psi_2(x,t)$ at the same location, 
and each QPC is described 
independently by a scattering matrix. 
\begin{figure}[htbp]
\includegraphics[width=7.5cm]{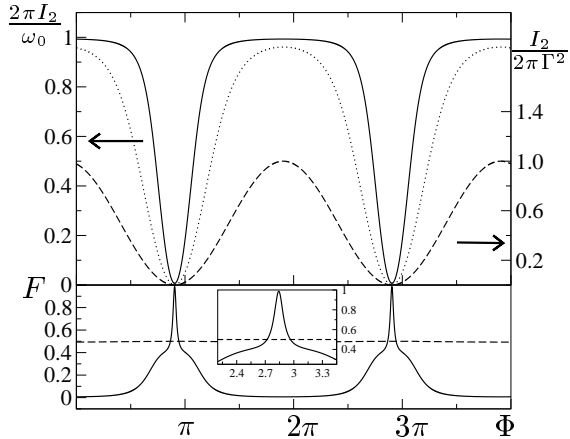}
\caption{Upper part: transmitted current $I_2$ as a function of the AB phase $\Phi$, for 2 QPCs with  $\nu=1/2$ and 
equal tunneling amplitudes $\Gamma$, with $\omega_0 \Delta/v_F= 0.3$, and $8 \pi \Gamma^2/\omega_0=0.01$ (dashed curve, right $y$ axis), 
$1$ (dotted curve, left axis), $100$ (full curve, left axis).  
Lower part: Fano factor $F=S_2/(2 e I_2)$ as a function of $\Phi$ for the same parameters, $8 \pi \Gamma^2/\omega_0=100$ (full curve),
 $0.01$ (dashed curve, extremely narrow peaks have been removed for the sake of clarity).
Inset: zoom on one of the narrow peaks of $F$ for $8 \pi \Gamma^2/\omega_0=100$.}
\label{fig:courant}
\end{figure}
This different behavior 
leads to dramatically different results for the interferences in the transmitted current $I_2$ 
as a function of the AB phase $\Phi$. Consider for simplicity the case where $|\Gamma_a| = |\Gamma_b| = \Gamma$.
Indeed, $\lambda=1$ merely ensures a maximum visibility for the interferences. 
In the IQHE, one has:
\begin{equation}
I_2|_{IQHE} = \omega_0 \, T (1-T) (1 + \cos(\omega_0 \Delta + \Phi))~,
\end{equation}
where $T = \Gamma^2 / (1+\Gamma^2/4)^2$ is the transmission of each QPC, and $\Phi$ is here $S B / \Phi_0 $ ($\Phi_0 = h c /e$).
 It shows that the maximum transmitted current in edge 2 is obtained
for $T=1/2$, while it goes to 0 for $T \to 0$ ($\Gamma \to 0$) or $T \to 1$ ($\Gamma \to 2$, which is the strong coupling limit in this case).
For all values of $T$, $I_2$ shows sinusoidal oscillations
as a function of $\Phi$.  Considering now the results for fractionaly charged excitations, Eqs.~(\ref{eq:currentsoluce}) and~(\ref{eq:gammaeff}),
one can distinguish two different regimes. First, the tunneling regime, corresponding to $\Gamma \to 0$.
We have then $\Gamma^2_{eff} = 2 \Gamma^2 (1 + \cos(\omega_0 \Delta + \Phi)) \ll 1$,  and we recover results similar to the non-interacting 
case in the tunneling limit: sinusoidal oscillations of the current $I_2$ as a function of the AB phase $\Phi$, with $I_2 \sim \Gamma^2 \ll 1$.
This is easily understood: when one keeps only the lowest order in $\Gamma$, the coupling between the two scattering events disappears and
the results obtained for non-interacting electrons are recovered. The opposite limit is obtained
when $\Gamma \to \infty$. As shown on Fig.~\ref{fig:courant}, the current $I_2$ is nearly constant, with the value $e \omega_0 /(2 \pi)$,
except near $\Phi= (2 n + 1) \pi$ where it shows narrow dips going to zero. For very large $\Gamma$, the width of the dips scales as 
$\sqrt{\omega_0 \Delta / v_F} < 1$. This means that for large $\Gamma$, all the incoming current gets transmitted to edge 2,
 except for special values of the AB phase $\Phi$ where destructive interference happens. 
Note that this is totally different from the non interacting electron case where 
the incoming current in edge 1 gets scattered to edge 2 at the first QPC, then gets mostly
scattered back in edge 1 at the second QPC. Because of electronic correlations, this picture is not valid in the FQHE, and the two QPCs must
be considered as a whole.

The noise also has unique features. Its analytic expression 
is identical to that of non-interacting electrons:
\begin{equation}
\label{eq:noise}
S_2 =  \frac{e^2}{2 \pi}  \, \int_{-\omega_0}^{\omega_0} \!\!\!\!\! d\omega \;
\left|t_{\omega}\right|^2 \left( 1 - \left|t_{\omega}\right|^2 \right)~,
\end{equation}
a mere consequence of the fact that the transmission in edge 2 is 
described by a scattering process for the refermionized field.
Here, however, the energy dependence of $\left|t_{\omega}\right|^2$
reflects the electronic correlations.    
The Fano factor $F\equiv S_2/(2 eI_2)$ is shown on the lower part of Fig.~\ref{fig:courant} for the two
regimes discussed above. In the tunneling regime, $F \simeq 1/2=\nu$:
the small current outgoing in edge 2 is carried by  quasiparticles of charge $\nu e$, 
and these can either tunnel at the first of at the second QPC. 
For arbitrary coupling, $F$ is a periodic function of flux:
in the regime of strong coupling, the lowering of the Fano factor 
is due to the factor $1 - |t_{\omega}|^2$ 
in Eq.~(\ref{eq:noise}), when the current is close 
to its maximal value. 
When destructive interference occurs (near $\Phi= (2p+1)\pi$, $p$ integer), 
$I_2$ is suppressed and a peculiar behavior appears.
The global shape of the Fano factor suggests a value of $1/2$ (sub-Poissonian)
in this region, although backscattering is strong. In the close 
vicinity of $\Phi=(2p+1)\pi$ there is a sharp peak, and 
the Fano factor reaches $1$ (Fig.~\ref{fig:courant}). 
For AB phases corresponding to this narrow peak, the noise is Poissonian, 
and the current is carried by pairs of quasiparticles of charge $\nu e$,
here electrons.   
This peak is in fact present for any value of $\Gamma$, 
but its width decreases with $\Gamma$ which makes it invisible in 
the small $\Gamma$ limit. For the large $\Gamma$ regime, 
and with  $\omega_0 \Delta/v_F \simeq 0.3$, 
this peak could be seen if currents of a few percent of the 
incoming Hall current can be measured experimentally.
All of the above results are robust up to $\omega_0 \Delta /v_F \simeq 1$:
beyond this value, the visibility of the current oscillations decreases rapidly and 
the ``Poissonian'' peak of the Fano factor is reduced. 

The chiral Luttinger liquid description is valid only for simple Laughlin 
fractions $\nu=1/(2p+1)$, not $\nu = 1/2$. We thus have to check that our results can
be observed with the experimentally accessible filling factors such as $\nu=1/3$.  
To this aim, we start with an imaginary time action formalism
and for simplicity we consider the case $\Gamma_a = \Gamma_b = \Gamma$. 
 Following Ref.~\cite{Furusaki-Nagaosa.1993}, we introduce the fields
$ \bar{\phi}(\omega),\widetilde{\phi}(\omega)  =  \left( \phi_{-}(x_a,\omega) \pm \phi_{-}(x_b,\omega) \right)/\sqrt{2}$.
These fields are the only degrees of freedom which are left after integration
of the quadratic part of the action associated with the Hamiltonian of Eq.~(\ref{eq:bosonHamiltonian}).
Assuming $\omega_0 \Delta /v_F < 1$, the effective action
reads:
\begin{multline}
\mathcal{S}= \frac{1}{2 \pi \beta} \left( \sum_{\omega} \frac{v_F}{\Delta} \left|\widetilde{\phi}(\omega) \right|^2
         + \frac{|\omega|}{2}  \left|\bar{\phi}(\omega) \right|^2 \right) \\
  + 4 \Gamma \int_0^{\hbar \beta} \!\!\!\! d\tau \; \cos(\sqrt{\nu} \, \bar{\phi}(\tau) + \Phi/2) 
\, \cos(\sqrt{\nu} \, \widetilde{\phi}(\tau) - \Phi/2) 
~.
\label{action_massive}
\end{multline}
where $\beta = 1/(k_B T)$. Note that in Eq. (\ref{action_massive}), the field $\widetilde{\phi}(\omega)$ is massive. 
For $\Delta <\hbar v_F/\Gamma$ we can neglect its fluctuations, so that the field $\widetilde{\phi}(\omega)$ is pinned to zero.
The leading corrections to this approximation are computed elsewhere \cite{the_return_of_thibaut}.
One is then left with the field $\bar{\phi}$ only. Shifting this field by $\Phi/2$, we 
get a new action:
\begin{equation}
\label{eq:BetheAction}
\mathcal{S}  = \frac{1}{4 \pi \hbar \beta} \sum_{\omega} |\omega| \left|\phi(\omega)\right|^2  +  4 \Gamma   \cos(\Phi/2) 
\int_0^{\hbar \beta} \!\!\!\! d\tau \; \cos(\sqrt{\nu} \, \phi(\tau))~. 
\end{equation}
This action is identical to the zero-mass limit of the Sine-Gordon model
and the problem can be solved exactly~\cite{Fendley-Ludwig.1995}.
The transmitted current $I_2$ follows the scaling: 
\begin{equation}
  I_2 = \frac{\nu}{2 \pi} \left( c_0 \Gamma \cos(\frac{\Phi}{2}) \right)^{\frac{1}{1-\nu}} 
  \, \mathcal{F}\left(\frac{\omega_0}{\nu \left(c_0 \Gamma \cos(\frac{\Phi}{2}) \right)^{\frac{1}{1-\nu}} }\right)
~,\end{equation}
with $c_0 =  4 \sqrt{2 \pi}$. 
$\mathcal{F}$ is the scaling function, with $\mathcal{F}(x) \sim x^{2 \nu -1}$ for $x \gg 1$ and $\mathcal{F}(x) = x$ for $x \ll 1$.
This proves that the 2 QPCs behave as 1 QPC with an effective coupling
 $\nu \left(4 \sqrt{2 \pi} \Gamma \cos(\Phi/2) \right)$ which is modulated by 
 the AB phase. The results previously obtained for $\nu=1/2$ 
can therefore be extended to describe Laughlin fractions, 
such as $\nu = 1/3$.
For $\nu=1/2$, $\mathcal{F}(x)$ is simply $\mbox{tan}^{-1}(x)$,
and the effective coupling is $\Gamma_{eff}^2 = 8 \pi \Gamma^2 (1+\cos \Phi)$. 
This is in agreement with Eq.~(\ref{eq:gammaeff}), since
by neglecting the massive field $\widetilde{\phi}$ we have supposed that $\Delta \to 0$.
 The results for the currrent $I_2$ when $\nu=1/3$ (not shown) are in precise correspondence 
 with those obtained for $\nu=1/2$ in the limit of $\Delta \to 0$.

Unusual features in the Fano factor (peaks near $\Phi = \pi, 3 \pi, \dots$) also need to be justified
for the filling factors $\nu = 1/(2 p +1)$. To this aim, we need to go beyond the infinite mass approximation:
the scattering term in Eq.~(\ref{eq:BetheAction}) is proportionnal to $\cos(\Phi/2)$ and is thus zero when $\Phi = \pi,3 \pi,\dots$.
As the transmission is very small in this region, we perform a perturbative developpement to get corrections:  
\begin{equation}
S_2 \simeq \frac{\Gamma^2 \Delta}{\hbar v_F} \sin^2(\Phi/2) \int_0^{\hbar \beta} \!\!\!\! d\tau \; \cos(2 \sqrt{\nu} \,\phi(\tau))~.
\end{equation}
Although this term is not relevant for $\nu > 1/4$, it gives the main contribution to $I_2$ and $S_2$ near the Fano factor peaks
where it is maximum. Because of the $\cos(2 \sqrt{\nu} \, \phi(\tau))$, it
implies the scattering of 2 excitations of charge $\nu e$ at once, and lead to an increase of the Fano factor from the expected $\nu$ value.
Narrow peaks in the Fano factor, near $\Phi=\pi,3 \pi,\dots$ are thus to be expected for filling factors $\nu=1/(2 p +1)$, as observed in the
calculations at $\nu=1/2$.

To conclude, we have provided the first non-perturbative treatment 
of the electronic analog of an optical interferometer operating 
with strongly correlated fermions. 
The most dramatic effect occurs when both QPCs are close to pinchoff
(large $\Gamma$), where the whole current exits in edge 2 (except at special values of the AB flux),
 contrary to the case of the IQHE and classical optics. The Fano factor is periodic in the 
AB phase. At strong pinchoff, at $\nu=1/2$,
the noise switches from sub-Poissonian to Poissonian
near the destructive AB interferences. 
Our predictions could be tested experimentally with the same 
``air bridge'' setup as in Ref.~\cite{Ji-Chung.2003}.
With $v_F \simeq 3. \, 10^5$ m/s~\cite{glattli_perso}, the important condition $\omega_0 \Delta / v_F \leq 1$ could be reached 
with state of the art techniques: 
temperature of  a few tens of  mK, $\Delta$ a few  $\mu\mathrm{m}$ and $V_0$ a few $\mu\mathrm{V}$.

Since the submission of this paper, a perturbative calculation on the same setup was presented in~\cite{Law-Feldman05}.

\end{document}